\documentclass[a4paper,fleqn,3p,onecolumn,12pt]{consistency-modmax_v5_elsarticle}

\newcommand{\versionnumber}{v5}

\journal{}  
\renewcommand{\today}{arXiv:1011.4258 (\versionnumber)}

\usepackage{graphicx}
\usepackage{bm}
\usepackage{mathrsfs}
\usepackage[latin1]{inputenc}
\usepackage{stmaryrd}
\usepackage{array}
\usepackage{psfrag}
\usepackage{dsfont}
\usepackage{epsfig}
\usepackage{titletoc}
\usepackage{float}   
\usepackage{wrapfig} 
\usepackage{amsmath}
\usepackage[latin1]{inputenc}
\usepackage{amsmath}\usepackage{amssymb,stmaryrd}
\usepackage{mathrsfs}
\usepackage{array}
\usepackage{graphicx}
\usepackage{psfrag}
\usepackage{dsfont}
\usepackage{epsfig}
\usepackage{cancel}

\def\be {\begin{equation}}
\def\ee {\end{equation}}
\def\ba {\begin{eqnarray}}
\def\ea {\end{eqnarray}}

\begin{document}

\begin{frontmatter}

\title{Consistency of isotropic modified Maxwell theory:\\
       Microcausality and unitarity}

\author{F.R. Klinkhamer\corref{cor1}}
\cortext[cor1]{Corresponding Author}\ead{frans.klinkhamer@kit.edu}
\author{M. Schreck} \ead{marco.schreck@kit.edu}
\address{Institute for Theoretical Physics, University of Karlsruhe,\\
         Karlsruhe Institute of Technology, 76128 Karlsruhe, Germany}

\begin{abstract}
The Lorentz-violating isotropic modified Maxwell theory minimally
coupled to standard Dirac theory is characterized by a single real
dimensionless parameter which is taken to vanish for the case of
the standard (Lorentz-invariant) theory.
A finite domain of positive and negative values of this Lorentz-violating
parameter is determined, in which microcausality and unitarity hold.
The main focus of this article is on isotropic modified Maxwell theory,
but similar results for an anisotropic nonbirefringent case are
presented in the appendix.
\end{abstract}
\end{frontmatter}

\setcounter{equation}{0}
\renewcommand{\theequation}{\arabic{section}.\arabic{equation}}
\section{Introduction}
\label{sec:Introduction}

There are two Lorentz-violating extensions of the standard theory
of photons~\cite{Heitler1954,JauchRohrlich1976,Veltman1994},
which are both gauge invariant and power-counting
renormalizable~\cite{ChadhaNielsen1983,ColladayKostelecky1998}.
The standard Lorentz-invariant Maxwell theory has a quadratic field
strength term ($F^2$) in the Lagrange density and the first
Lorentz-violating extension adds a CPT-odd Chern--Simons-type term
($m_\text{CS}\,\widehat{k}\,A\,F$, with a fixed  normalized
``four-vector'' $\widehat{k}^\mu$ and mass scale $m_\text{CS}$).
The second Lorentz-violating extension adds another $F^2$ term,
which has different contractions than those of the standard Maxwell term.

The consistency of the CPT-violating Maxwell--Chern--Simons (MCS)
theory~\cite{Carroll-etal1990}
has been studied in Ref.~\cite{AdamKlinkhamer2001}
and the result is that certain choices of the parameters
(specifically, timelike $\widehat{k}^\mu$) lead
to violation of microcausality and/or unitarity. The concern now is
the consistency of the CPT-invariant modified Maxwell theory,
in particular, the theory restricted to the isotropic sector.

The isotropic modified Maxwell theory is described by a single real
dimensionless parameter $\widetilde{\kappa}_{\mathrm{tr}}$
and is, therefore, one of the simplest possible examples of
a CPT--even Lorentz-violating theory of photons.
The standard Lorentz-invariant Maxwell theory has
$\widetilde{\kappa}_{\mathrm{tr}}=0$.
Positive values of $\widetilde{\kappa}_{\mathrm{tr}}$
have been derived from an underlying small-scale structure of
spacetime in the long-wavelength limit of (standard)
photons~\cite{KlinkhamerRupp2004,BernadotteKlinkhamer2007},
so that isotropic modified Maxwell theory with small enough
positive $\widetilde{\kappa}_{\mathrm{tr}}$ can be expected
to be consistent. But the consistency of isotropic modified
Maxwell theory for \emph{negative} values
of $\widetilde{\kappa}_{\mathrm{tr}}$
is an entirely open question.
In fact, there are only partial results
for $\widetilde{\kappa}_{\mathrm{tr}}\geq 0$
in the literature~\cite{Casana-etal2009,Casana-etal2010},
which makes it worthwhile to give a
more or less comprehensive analysis of the isotropic case.
In addition, there have been many experiments
(selected references will be given in Section~\ref{sec:Discussion-outlook})
which provide upper and lower bounds
on the parameter $\widetilde{\kappa}_{\mathrm{tr}}$, simply
assuming the theory to be consistent and open to experimental
verification.

The outline of this article is as follows.
A brief discussion of isotropic modified Maxwell theory is given in
Sections~\ref{subsec:Action-nonbirefringent-Ansatz}
and \ref{subsec:Restriction-isotropic}.
The pure-photon theory is extended by the introduction of a
minimal coupling of this photon to a charged Dirac particle
in Section~\ref{subsec:Coupling-to-standard-Dirac-particles}.
In short, the theory considered is a particular modification
of standard quantum electrodynamics, with
a modified kinetic term of the photon in the action.
The corresponding gauge-field propagator in Feynman gauge
is presented in Section~\ref{sec:Propagator-isotropic}.
Microcausality (i.e., commutation of electric and magnetic field
operators with certain spacelike separations)
is established in Section~\ref{sec:Microcausality-isotropic},
together with the global causality of the theory
(e.g., absence of closed timelike loops).
Reflection positivity of the Euclidean gauge-field propagator
is demonstrated in Sections~\ref{subsec:Reflection-positivity-general}
and \ref{subsec:Reflection-positivity-isotropic}.
The unitarity of the interacting theory is checked by
the direct evaluation of the optical theorem for one particular process
in Section~\ref{subsec:Optical-theorem-isotropic}.
Concluding remarks are presented in Section~\ref{sec:Discussion-outlook}.
The results for an anisotropic nonbirefringent
case are given in Appendix~\ref{sec:Parity-even-anisotropic}.

\setcounter{equation}{0}
\section{Isotropic modified Maxwell theory}
\label{sec:Isotropic-ModMaxtheory}

\subsection{Action and nonbirefringent Ansatz}
\label{subsec:Action-nonbirefringent-Ansatz}

In this article, we consider modified Maxwell
theory~\cite{ChadhaNielsen1983,ColladayKostelecky1998,KosteleckyMewes2002}
which has an action given by
\begin{subequations}\label{eq:action-modified-maxwell-theory}
\begin{eqnarray}
S_{\mathrm{modMax}}&=&\int_{\mathbb{R}^4}\mathrm{d}^4x\;
\mathcal{L}_\text{modMax}(x)\,,\\[2mm]
\mathcal{L}_\text{modMax}(x)&=& -\frac{1}{4}\,
\eta^{\mu\rho}\,\eta^{\nu\sigma}\,F_{\mu\nu}(x)F_{\rho\sigma}(x)
-\frac{1}{4}\,
\kappa^{\mu\nu\varrho\sigma}\,F_{\mu\nu}(x)F_{\varrho\sigma}(x)\,,
\label{eq:L-modified-maxwell-theory}
\end{eqnarray}
\end{subequations}
where $F_{\mu\nu}(x)\equiv\partial_{\mu}A_{\nu}(x)-\partial_{\nu}A_{\mu}(x)$ is
the field strength tensor of the $U(1)$ gauge field $A_{\mu}(x)$.
The photons propagate over a flat Minkowski spacetime
with global Cartesian coordinates
$(x^\mu)$ $=$ $(x^0,\boldsymbol{x})$ $=$ $(c\,t,x^1,x^2,x^3)$
and metric $g_{\mu\nu}(x)$ $=$ $\eta_{\mu\nu}$ $\equiv$
$\text{diag}\,(1,\, -1,\, -1,\, -1)\,$.
The fixed spacetime-independent background
field $\kappa^{\mu\nu\varrho\sigma}$ in the second term of
\eqref{eq:L-modified-maxwell-theory} manifestly breaks Lorentz
invariance.

If $\kappa^{\mu\nu\varrho\sigma}$ is taken to have a vanishing double
trace, $\kappa^{\mu\nu}_{\phantom{\mu\nu}\mu\nu}=0$, and to obey the same
symmetries as the Riemann curvature tensor, the number of independent
Lorentz-violating parameters is 19.
To leading order,  birefringence is controlled by
10 of these 19 parameters. We restrict our considerations to the
nonbirefringent sector with 9 parameters, which is parameterized by the
following \textit{Ansatz}~\cite{BaileyKostelecky2004}:
\begin{equation}\label{eq:nonbirefringent-Ansatz}
\kappa^{\mu\nu\varrho\sigma}=
\frac{1}{2}\,\Big(
 \eta^{\mu\varrho}\,\widetilde{\kappa}^{\nu\sigma}
-\eta^{\mu\sigma}\,\widetilde{\kappa}^{\nu\varrho}
-\eta^{\nu\varrho}\,\widetilde{\kappa}^{\mu\sigma}
+\eta^{\nu\sigma}\,\widetilde{\kappa}^{\mu\varrho}\Big)\,.
\end{equation}
The constant $4\times 4$ matrix $\widetilde{\kappa}^{\mu\nu}$ is
symmetric and traceless. Here and in the following, natural units
are used with $\hbar=c=1$, where $c$ corresponds to the maximal
attainable velocity of a standard Dirac particle
(see Section~\ref{subsec:Coupling-to-standard-Dirac-particles}).

\subsection{Restriction to the isotropic case}
\label{subsec:Restriction-isotropic}

Next, restrict the nonbirefringent modified Maxwell theory to the
isotropic sector which is characterized by a purely timelike
normalized four-vector
$\xi^{\mu}$ in a preferred reference frame
and a single real dimensionless parameter
$\widetilde{\kappa}_{\mathrm{tr}}$\,:
\begin{subequations}\label{eq:definition-isotropic-case}
\begin{eqnarray}
\widetilde{\kappa}^{\mu\nu}
&=&2\,\widetilde{\kappa}_{\mathrm{tr}}
\left(\xi^{\mu}\xi^{\nu}-
\frac{1}{4}\, \xi^{\lambda}\xi_{\lambda}\,\eta^{\mu\nu}
\right)\,,\\[2mm]
(\xi^{\mu})&=&(1,\,0,\,0,\,0)\,,
\label{eq:definition-isotropic-case-xi}\\[2mm]
(\widetilde{\kappa}^{\mu\nu})
&=&\frac{3}{2}\;
\widetilde{\kappa}_{\mathrm{tr}}\,\mathrm{diag}
\left(1,\,\frac{1}{3},\,\frac{1}{3},\,\frac{1}{3}\right)\,.
\label{eq:definition-isotropic-case-widetildekappamunu}
\end{eqnarray}
\end{subequations}
 From \eqref{eq:action-modified-maxwell-theory}--\eqref{eq:definition-isotropic-case},
the Lagrange density becomes in terms of the standard electric field
$E^i\equiv F^{i0}$ and
magnetic field $B^i\equiv (1/2)\, \epsilon_{ijk}\,F^{jk}$:
\begin{equation}\label{eq:L-modMax-isotropic}
\mathcal{L}_\text{modMax}^\text{isotropic}
\big[c,\widetilde{\kappa}_\text{tr}\big](x)
= \frac{1}{2}\,\Big(
(1+\widetilde{\kappa}_\text{tr})\,|\mathbf{E}(x)|^2 -
(1-\widetilde{\kappa}_\text{tr})\,|\mathbf{B}(x)|^2
\Big)\,,
\end{equation}
where the dependence on the fundamental constants
$c$ and $\widetilde{\kappa}_\text{tr}$ has been made explicit
on the left-hand side, which will be useful for the discussion of
unitarity later.

The field equations of modified Maxwell theory in momentum space,
\begin{equation}
M^{\mu\nu}A_{\nu}=0 \,,\quad
M^{\mu\nu}\equiv
k^{\lambda}k_{\lambda}\,\eta^{\mu\nu}-k^{\mu}k^{\nu}
-2\,\kappa^{\mu\rho\sigma\nu}\,k_{\rho}k_{\sigma}\,,
\label{eq:field-equations-modified-maxwell-theory}
\end{equation}
give the following dispersion relation for the isotropic case:
\begin{equation}
\omega(\mathbf{k})=\mathcal{B}\,|\mathbf{k}|\,,\quad
\mathcal{B}\equiv\sqrt{\frac{1-\widetilde{\kappa}_{\mathrm{tr}}}{1+\widetilde{\kappa}_{\mathrm{tr}}}}\,,
\label{eq:dispersion-isotropic}
\end{equation}
in terms of the norm of the momentum three-vector $\mathbf{k}$,
explicitly defined
by $|\mathbf{k}| \equiv \big((k_1)^2+(k_2)^2+(k_3)^2\big)^{1/2}$.
The additional constant $\mathcal{A}\equiv\mathcal{B}^{-1}$ has been
used in Ref.~\cite{KlinkhamerSchreck2008}, but, in the present article,
we prefer to employ only the constant $\mathcal{B}$.

The dispersion relation \eqref{eq:dispersion-isotropic}
yields the following phase velocity of electromagnetic waves:
\begin{equation}
v_{\mathrm{ph}}\equiv \frac{\omega(\mathbf{k})}{|\mathbf{k}|}
               =\mathcal{B}\,.
\label{eq:phase-velocity-isotropic}
\end{equation}
This phase velocity equals the group velocity,
\begin{equation}
v_{\mathrm{gr}}\equiv
\left|\frac{\partial\, \omega(\mathbf{k})}{\partial\, \mathbf{k}}\right|
=\mathcal{B}\,,
\end{equation}
which implies that the shape of a wave package does not change
with time. From the modified dispersion law
\eqref{eq:dispersion-isotropic}, it is clear that the vacuum
behaves like an effective medium with a refraction index
\begin{equation}
n\equiv \frac{|\mathbf{k}|}{\omega(\mathbf{k})}=\mathcal{B}^{-1}\,,
\end{equation}
which is frequency independent because $\mathcal{B}^{-1}$
is a constant. This particular Lorentz-violating theory does
not display dispersion effects of photons propagating in vacuum.

Unless stated otherwise, we henceforth restrict
$\widetilde{\kappa}_{\mathrm{tr}}$ to the following half-open interval:
\begin{equation}\label{eq:kappatildetrace-domain}
\widetilde{\kappa}_{\mathrm{tr}}\in I\,,\quad I\equiv (-1,1]\,,
\end{equation}
since for $\widetilde{\kappa}_{\mathrm{tr}}\notin I$ the
frequency \eqref{eq:dispersion-isotropic} becomes complex
(signaling unstable behavior).
The front velocity, which corresponds to the velocity of the
high-frequency forerunners of electromagnetic waves~\cite{Brillouin1960},
is given by
\begin{equation}
v_{\mathrm{fr}}\equiv\lim_{k\mapsto\infty} v_{\mathrm{ph}}= \mathcal{B}\,,
\end{equation}
and is seen to be equal to both the phase and group velocity. For
$\widetilde{\kappa}_{\mathrm{tr}}<0$, the front velocity of light exceeds the
maximum attainable velocity of the standard matter particles,
$v_{\mathrm{fr}}> c\equiv 1$. This alerts us to the issue of causality,
which
will be discussed in Section~\ref{sec:Microcausality-isotropic}.

\subsection{Coupling to matter: Modified QED}
\label{subsec:Coupling-to-standard-Dirac-particles}

For the coupling of the photon to matter, take the minimal coupling
to a standard (Lorentz-invariant)
spin-$\textstyle{\frac{1}{2}}$ Dirac particle with electric charge $e$
and mass $M$. That is, the theory considered is a particular
deformation of quantum electrodynamics
(QED)~\cite{Heitler1954,JauchRohrlich1976,Veltman1994}
given by the following action:
\begin{equation}\label{eq:action-isotropic-modQED} \hspace*{0mm}
S_\text{modQED}^\text{isotropic}\big[c,\widetilde{\kappa}_\text{tr},e,M\big] =
S_\text{modMax}^\text{isotropic}\big[c,\widetilde{\kappa}_\text{tr}\big] +
S^\text{ }_\text{Dirac}\big[c,e,M\big] \,,
\end{equation}
with the modified-Maxwell term
\eqref{eq:action-modified-maxwell-theory}--\eqref{eq:L-modMax-isotropic}
for the gauge field $A_\mu(x)$
and the standard Dirac term for the spinor field $\psi(x)$,
\begin{equation}\label{eq:standDirac-action}
S^\text{ }_\text{Dirac}\big[c,e,M\big] =
\int_{\mathbb{R}^4} \mathrm{d}^4 x \; \overline\psi(x) \Big(
\gamma^\mu \big(\mathrm{i}\,\partial_\mu -e A_\mu(x) \big) -M\Big) \psi(x)\,,
\end{equation}
with standard Dirac matrices $\gamma^\mu$ corresponding to the Minkowski
metric $\eta^{\mu\nu}$. As mentioned before, the fundamental constant $c$
may be operationally defined as the maximum attainable velocity of the
Dirac particle. For further discussion
on Lorentz violation and the role of different particle species, see, e.g.,
Refs.~\cite{ColemanGlashow1998,Freidel-etal2003,LammerzahlHehl2004}
and references therein.

\setcounter{equation}{0}
\section{Polarization sum and propagator}
\label{sec:Propagator-isotropic}

The polarization sum can be computed by solving the field equations
\eqref{eq:field-equations-modified-maxwell-theory}
for appropriate coupling constants $\kappa^{\mu\rho\sigma\nu}$,
while respecting the normalization condition
\begin{equation}
\langle\mathbf{k},\lambda|\; :\hspace{-0cm}P^0
\hspace{-0cm}: \;|\mathbf{k},\lambda\rangle=
\langle\mathbf{k},\lambda|\;\int\mathrm{d}^3x\,:
\hspace{-0cm}T^{00}\hspace{-0cm}: \;\;|\mathbf{k},\lambda\rangle
\equiv \omega(\mathbf{k})\,,
\label{eq:normalization-polarization-vectors}
\end{equation}
where, as usual, the pair of colons
stands for the normal ordering of
the enclosed operators and $|\mathbf{k},\lambda\rangle$
denotes a photon state with momentum three-vector $\mathbf{k}$ and
polarization label $\lambda$. The $T^{00}$ component of the
energy-momentum tensor can be cast in the following
form~\cite{ColladayKostelecky1998}:
\begin{eqnarray}\label{eq:t00-component-modified-maxwell-theory}
T^{00}&=&\frac{1}{2}\,\Big(|\mathbf{E}|^2+|\mathbf{B}|^2\Big)
-\kappa^{0j0k}E^jE^k
+\frac{1}{4}\,\kappa^{jklm}\varepsilon^{jkp}\varepsilon^{lmq}\,B^pB^q
\nonumber\\[1mm]
&=&
\frac{1}{2}\,\Big(
(1+\widetilde{\kappa}_\text{tr})\,|\mathbf{E}|^2 +
(1-\widetilde{\kappa}_\text{tr})\,|\mathbf{B}|^2
\Big)\,,
\end{eqnarray}
with the three-dimensional totally antisymmetric Levi-Civita symbol
$\varepsilon^{ijk}$ and the electric and magnetic field components $E^i$
and $B^j$ defined in the lines above \eqref{eq:L-modMax-isotropic}.
The final expression  in \eqref{eq:t00-component-modified-maxwell-theory}
makes clear that, for $|\widetilde{\kappa}_\text{tr}|>1$, the theory
suffers from unavoidable instabilities if the coupling
to matter \eqref{eq:action-isotropic-modQED} is taken into account
(see, e.g., Ref.~\cite{KosteleckyLehnert2000} for a general discussion
of the energy-positivity condition).

Returning to the parameter domain \eqref{eq:kappatildetrace-domain},
the solution of the field equations and the resulting
energy-momentum tensor component
\eqref{eq:t00-component-modified-maxwell-theory}
give the following expression for the polarization sum:
\begin{eqnarray}
\Pi^{\mu\nu}&\equiv&\sum_{\lambda=1,\, 2}
\overline{(\varepsilon^{(\lambda)})}^{\,\mu}
\;(\varepsilon^{(\lambda)})^{\nu}
\nonumber\\[1mm]
&=&
\frac{1}{1+\widetilde{\kappa}_{\mathrm{tr}}}\Biggl(-\eta^{\mu\nu}
\,-\frac{1}{|\mathbf{k}|^2}\,k^{\mu}k^{\nu}
+\frac{\mathcal{B}}{|\mathbf{k}|}\,
\big(k^{\mu}\xi^{\nu}+\xi^{\mu}k^{\nu}\big)
+\frac{2\,\widetilde{\kappa}_{\mathrm{tr}}}{1+\widetilde{\kappa}_{\mathrm{tr}}}
\,\xi^{\mu}\xi^{\nu}\Bigr. \Bigl.\Biggr)\,,
\label{eq:polarization-sum-isotropic}
\end{eqnarray}
where the sum runs over the two physical polarizations $\lambda\in \{1,2\}$,
with the polarization vectors $\varepsilon^{(1)}$ and $\varepsilon^{(2)}$
being orthogonal to the momentum three-vector $\mathbf{k}$.
Expression \eqref{eq:polarization-sum-isotropic}
for $\widetilde{\kappa}_{\mathrm{tr}}=0$
reproduces the standard result~\cite{Veltman1994}.
For later use, also the expression is given if terms with
$k^\mu$ or $k^\nu$ are removed:
\begin{eqnarray}
\Pi^{\mu\nu}\;\Big|^\text{truncated} &=&
\frac{1}{1+\widetilde{\kappa}_{\mathrm{tr}}}\Biggl(-\eta^{\mu\nu}
+\frac{2\,\widetilde{\kappa}_{\mathrm{tr}}}{1+\widetilde{\kappa}_{\mathrm{tr}}}
\,\xi^{\mu}\xi^{\nu}\Bigr. \Bigl.\Biggr)\,.
\label{eq:polarization-sum-isotropic-truncated}
\end{eqnarray}

For the study of causality and unitarity in the quantum theory,
the gauge-field propagator $G_{\mu\nu}$ is needed.
Recall that $G_{\mu\nu}$ is the Green's
function of the free field equations of the gauge potential
in momentum space, which requires a particular gauge fixing.

The Feynman gauge~\cite{Veltman1994,ItzyksonZuber1980,PeskinSchroeder1995},
for example, is implemented by the following gauge-fixing condition:
\begin{equation}
\mathcal{L}_{\mathrm{gf}}(x)=
-\frac{1}{2}\big(\partial_{\mu}\,A^{\mu}(x)\big)^2\,.
\label{eq:gauge-fixing-feynman}
\end{equation}
It has been proven in Ref.~\cite{BialynickiBirula1970} that the Feynman gauge
in standard QED is equivalent to the Coulomb gauge, if one computes transition
probabilities, i.e., matrix-element squares. The Coulomb gauge is well-defined
in the sense that any gauge field configuration can be brought into that form.
Since isotropic modified Maxwell theory is a deformation of standard QED,
we expect this to be valid also in our case, at least, for a small enough
deformation parameter $|\widetilde{\kappa}_\text{tr}|$.

The differential operator $(G^{-1})^{\mu\nu}$ in
Feynman gauge reads
\begin{equation}
(G^{-1})^{\mu\nu}=\eta^{\mu\nu}\partial^2
-2\, \kappa^{\mu\varrho\sigma\nu}\partial_{\varrho}\partial_{\sigma}\,,
\end{equation}
and the gauge-field propagator is
\begin{align}
G_{\nu\lambda}\,\big|^{\mathrm{Feynman}}
&=-\mathrm{i}\,\Big\{a\,\eta_{\nu\lambda}+b\,k_{\nu}k_{\lambda}+c\,\xi_{\nu}\xi_{\lambda}
+d\,\big(k_{\nu}\xi_{\lambda}+\xi_{\nu}k_{\lambda})
\Big\}\,K\,,
\label{eq:propagator-result-isotropic}
\end{align}
with scalar propagator $K$ and coefficient functions
$a$, $b$, $c$, $d$ given by
\begin{equation}
K=\frac{1}{\left(1-\widetilde{\kappa}_{\mathrm{tr}}\,\xi^2\right)k^2
+2\,\widetilde{\kappa}_{\mathrm{tr}}(k\cdot \xi)^2}\,,
\label{eq:propagator-result-isotropic-scalar-part}
\end{equation}
\begin{subequations}\label{eq:propagator-result-isotropic-coefficients}
\begin{equation}
a=1\,,
\end{equation}
\begin{align}
b=\frac{\widetilde{\kappa}_{\mathrm{tr}}\;
\xi^2}{1+\widetilde{\kappa}_{\mathrm{tr}}\,\xi^2}\;
\frac{1}{k^4}\Big(-\left(1+\widetilde{\kappa}_{\mathrm{tr}}\,\xi^2\right)k^2+2\,
\widetilde{\kappa}_{\mathrm{tr}}(\xi\cdot k)^2\Big)\,,
\end{align}
\begin{equation}\label{eq:propagator-result-isotropic-c-coeff}
c=-\frac{2\,\widetilde{\kappa}_{\mathrm{tr}}}{1+\widetilde{\kappa}_{\mathrm{tr}}\,\xi^2}\,,
\end{equation}
\begin{equation}
d=\frac{2\,\widetilde{\kappa}_{\mathrm{tr}}}{1+\widetilde{\kappa}_{\mathrm{tr}}\,\xi^2}
\;
\frac{\xi\cdot k}{k^2}\,.
\end{equation}
\end{subequations}

Casana et al. \cite{Casana-etal2010} have also obtained the gauge-field propagator in Feynman gauge from a more general \textit{Ansatz}
given by their Eq.~(28) with two four-vectors $U^{\mu}$ and $V^{\mu}$.
Their result for $(U^{\mu})=(V^{\mu})=
(\sqrt{2\,\widetilde{\kappa}_{\mathrm{tr}}},\,0,\,0,\,0)$
agrees with ours.

\setcounter{equation}{0}
\section{Microcausality}
\label{sec:Microcausality-isotropic}

\subsection{Commutators of gauge potentials and physical fields}
\label{subsec:Commutator-gauge-potential}

The notion of microcausality can be condensed to the statement
that the commutator of two field operators $\Phi(x')$ and $\Phi(x'')$
must vanish for spacelike separations, specifically,
$[\Phi(x'),\Phi(x'')]=0$ for $(x'-x'')^2<0$. This assures that information
can only propagate along or inside null-cones. Translation invariance of
the modified Maxwell theory implies the following structure of
the gauge-field commutator:
\begin{equation}
K_{\mu\nu}(x',x'')\equiv
[A_{\mu}(x'),A_{\nu}(x'')]=[A_{\mu}(x'-x''),A_{\nu}(0)]=
[A_{\mu}(x),A_{\nu}(0)]\,,
\end{equation}
for $x_{\mu}\equiv x'_{\mu}-x''_{\mu}$.
The corresponding result in momentum space must be of the form
\begin{equation}
K_{\mu\nu}(k)=
\Xi_{\mu\nu}(k^0,\mathbf{k})\,
\Big(\mathrm{i}D(k)\Big)\,,
\end{equation}
where
$\Xi_{\mu\nu}(k)$ respects the tensor structure of the commutator
and $D(k)$ is a scalar commutator function.

The commutator $K_{\mu\nu}(k)$ can be computed
either directly by Fourier decomposition of the gauge potential
in positive and negative frequency parts or by extraction from the
Feynman propagator. Both methods yield the same result:
\begin{subequations}\label{eq:commutator-gaugepotential-isotropic}
\begin{eqnarray}
\Xi_{\mu\nu}&=&(1+\widetilde{\kappa}_{\mathrm{tr}})\,\Pi_{\mu\nu}\,,
\label{eq:commutator-gaugepotential-isotropic-Xi}\\[1mm]
D(k)^{-1}&=& (1+\widetilde{\kappa}_{\mathrm{tr}})\,k_0^2
             -(1-\widetilde{\kappa}_{\mathrm{tr}})\,|\mathbf{k}|^2\,,
\label{eq:commutator-gaugepotential-isotropic-D}
\end{eqnarray}
\end{subequations}
where $\Pi_{\mu\nu}$ is the polarization sum
\eqref{eq:polarization-sum-isotropic}.
In fact, \eqref{eq:polarization-sum-isotropic}
gives  $\Xi_{00}=\Xi_{0m}=\Xi_{m0}=0$ for $m\in \{1,2,3\}$, so that
only certain purely spatial components of $K_{\mu\nu}$ may
be nonzero.

By using \eqref{eq:commutator-gaugepotential-isotropic},
the momentum-space commutators of the electric and magnetic fields
can be computed, which are then transformed to configuration space.
The resulting commutators are
\begin{subequations}\label{eq:commutator-ee-eb-bb}
\begin{equation}
[ E_i(x),E_j(0) ]=
(\partial_0^2\,\delta_{ij}-\mathcal{B}^{2}\,\partial_i\partial_j)\,
\big(\mathrm{i}D(x)\big)\,,
\label{eq:commutator-ee}
\end{equation}
\begin{equation}
[ E_{i}(x),B_{j}(0) ]=
\varepsilon_{ijk}\, \partial_0\partial_k\,\big(\mathrm{i}D(x)\big)\,,
\label{eq:commutator-eb}
\end{equation}
\begin{equation}
[ B_{i}(x),B_{j}(0) ]=
(\nabla^2\,\delta_{ij}-\partial_i\partial_j)\,\big(\mathrm{i}D(x)\big)\,,
\label{eq:commutator-bb}
\end{equation}
\end{subequations}
where $\mathcal{B}$ has been defined in \eqref{eq:dispersion-isotropic}.
The scalar commutator function in configuration space can be written as
follows
\begin{equation}
D(x)=\oint_{C} \;\frac{\mathrm{d}k_0}{2\pi}\, \int
\frac{\mathrm{d}^3k}{(2\pi)^{3}\,}\,
\frac{1}{\left(1+\widetilde{\kappa}_{\mathrm{tr}}\right)k_0^2-\left(1-\widetilde{\kappa}_{\mathrm{tr}}\right)|\mathbf{k}|^2}
\;\exp\big(\mathrm{i}\,k_{0}\,x_{0}
+\mathrm{i}\,\mathbf{k}\cdot \mathbf{x}\big)\,,
\label{eq:commutator-function-configuration-space}
\end{equation}
where the poles are circled in the counterclockwise direction
along a contour $C$. The evaluation of the four-dimensional integral
\eqref{eq:commutator-function-configuration-space} leads to the following
expression:
\begin{equation}
D(x)=-\frac{1}{2\pi\sqrt{1-\widetilde{\kappa}_{\mathrm{tr}}^2}}\;
\mathrm{sgn}(\widetilde{x}_0)\,
 \delta\big((\widetilde{x}_0)^2-|\mathbf{x}|^2\big)\,,\quad
\widetilde{x}_0\equiv \mathcal{B}\,x_0\,,
\label{eq:commutator-function-configuration-space-evaluated}
\end{equation}
with the sign function
\begin{equation}
\mathrm{sgn}(x)=\left\{\begin{array}{rcl}
1 & \text{for} & x>0 \\
0 & \text{for} & x=0 \\
-1 & \text{for} & x<0 \\
\end{array}
\right.\,.
\end{equation}
The overall minus sign in \eqref{eq:commutator-function-configuration-space-evaluated}
has its origin in the definition of the commutator function
\eqref{eq:commutator-function-configuration-space}. In this definition,
the first term in the argument of the exponential function
enters with a plus sign.
This convention has been chosen to conform with Ref.~\cite{AdamKlinkhamer2001}
and is different from the one used in, for example,
Appendix~A1 of Ref.~\cite{JauchRohrlich1976}.
The commutators of the physical electric and magnetic
fields in \eqref{eq:commutator-ee-eb-bb} are, of course,
independent of this convention.
For $\widetilde{\kappa}_{\mathrm{tr}}\mapsto 0$, these commutators
are equal to those of standard
QED, first obtained by Jordan and Pauli~\cite{JordanPauli1928,Heitler1954}.

According to \eqref{eq:commutator-function-configuration-space-evaluated},
the commutators \eqref{eq:commutator-ee-eb-bb}
vanish if the relative distance in Minkowski spacetime lies outside
the modified null-cone,
\begin{equation}
(\widetilde{x}_0)^2-|\mathbf{x}|^2=0\,.
\label{eq:minkowski-distance-square-modified}
\end{equation}
The same result holds in standard QED with the standard null-cone,
$x_0^2-|\mathbf{x}|^2=0$.
Modified Maxwell theory only has a different opening angle
of the null-cone for the case of a nonzero value of the
Lorentz-violating parameter $\widetilde{\kappa}_{\mathrm{tr}}$.

Different from the commutators of Maxwell-Chern-Simons (MCS)
theory with a spacelike parameter~\cite{AdamKlinkhamer2001},
the commutators
\eqref{eq:commutator-ee-eb-bb}--\eqref{eq:commutator-function-configuration-space-evaluated}
vanish everywhere except on the null-cone, since the pure-photon sector of
modified Maxwell theory is scale-invariant. The spacelike MCS theory,
on the other hand, is characterized by a mass scale,
called $m_\text{CS}$ in Section~\ref{sec:Introduction}, which leads to
nonvanishing commutators both on and inside the null-cone.

Returning to the isotropic modified Maxwell theory, consider,
now, the interaction of photons and charged matter particles
as given by the modified QED action \eqref{eq:action-isotropic-modQED}
and take, for simplicity, a vanishing mass for the Dirac particle,
$M=0$. Then, the photon has a null-cone
\eqref{eq:minkowski-distance-square-modified}
and the Dirac particle a different one given by $x_0^2-|\mathbf{x}|^2=0$.
Intuitively, there are no causality problems to be
expected from having these two different null-cones.
There may, of course, be nonstandard interaction processes,
for example, vacuum Cherenkov radiation
for $\widetilde{\kappa}_{\mathrm{tr}}>0$
and photon decay for $\widetilde{\kappa}_{\mathrm{tr}}<0$
(see Ref.~\cite{KlinkhamerSchreck2008} for detailed calculations).

\subsection{Wick rotation}
\label{subsec:Wick-rotation}

For the analytic properties of the gauge-field
propagator \eqref{eq:propagator-result-isotropic},
the behavior of the propagator
pole structure under Wick rotation is an important issue.
For $\widetilde{\kappa}_{\mathrm{tr}}$
in the domain \eqref{eq:kappatildetrace-domain},
the scalar part \eqref{eq:propagator-result-isotropic-scalar-part}
of the propagator
shows that by performing a Wick rotation
the poles of the full propagator do not lie within the
integration contour and that the Wick-rotated axes do not
cross any poles.

But these properties hold only for
$\widetilde{\kappa}_{\mathrm{tr}}\in (-1,1]$.
A Wick rotation $k_4=-\mathrm{i}k^0$ from  Minkowski spacetime
to Euclidean space, for example, maps poles with positive real part
in Minkowski spacetime to poles with negative
imaginary part in Euclidean space or poles with positive imaginary part to
poles with negative real part. Hence, an analytic continuation of the
propagator from Minkowski spacetime to Euclidean space, or
\textit{vice versa}, is possible with the Wick rotation. The
gauge-field propagator \eqref{eq:propagator-result-isotropic}
is thus well-behaved for the above mentioned parameter domain.
For $\widetilde{\kappa}_{\mathrm{tr}}\notin (-1,1]$, however, the
$k_0$ poles in Minkowski spacetime lie on the imaginary axis,
which implies that the corresponding
energy becomes imaginary for this parameter domain.

\subsection{Effective metric}
\label{subsec:Effective-metric}

Following up on earlier work about the coupling of
Lorentz-violating theories to gravity~\cite{Kostelecky2004},
it has been shown in Refs.~\cite{BetschartKK2008,KantKS2009}
that the action of isotropic modified Maxwell
theory from  \eqref{eq:action-modified-maxwell-theory} can be cast in
the following form:
\begin{equation}
S^\text{isotropic}_{\mathrm{modMax}}=
-\,\left(1-\widetilde{\kappa}_{\mathrm{tr}}\right)\,
\int_{\mathbb{R}_4} \mathrm{d}^4x\;
\frac{1}{4}\;
\widetilde{\eta}^{\mu\varrho}\,\widetilde{\eta}^{\nu\sigma}\,
F_{\mu\nu}F_{\varrho\sigma}\,,
\end{equation}
with an effective metric
\begin{equation}
\widetilde{\eta}^{\mu\nu}=
\eta^{\mu\nu}+\frac{2\,\widetilde{\kappa}_{\mathrm{tr}}}
{1-\widetilde{\kappa}_{\mathrm{tr}}}\;
\xi^{\mu}\xi^{\nu}\,,
\label{eq:effective-metric-isotropic}
\end{equation}
for $\xi^\mu$ from \eqref{eq:definition-isotropic-case-xi}.
The existence of such an effective metric has interesting implications.

First, recall that a spacetime manifold $M_{4}$
is said to be ``stably causal''
if and only if there exists a Lorentzian metric $g_{\mu\nu}(x)$
and a scalar function $\theta(x)$, defined everywhere on $M_{4}$,
so that $\nabla_{\mu}\theta\neq 0$ and
$g^{\mu\nu}\,(\nabla_{\mu}\theta)(\nabla_{\nu}\theta)>0$.
If a spacetime manifold is stably causal,
it does not contain closed timelike or lightlike
curves (cf. Section~6.4 of Ref.~\cite{HawkingEllis1973}).

For the isotropic case of modified Maxwell theory
defined over Minkowski spacetime with standard global spacetime
coordinates as given below \eqref{eq:action-modified-maxwell-theory},
we can choose the globally defined scalar
function $\theta(x)$ to be given by the time coordinate $t$.
Then, $(\nabla_{\mu}t)=(1,0,0,0)\neq 0$
and the effective metric \eqref{eq:effective-metric-isotropic} gives:
\begin{equation}
\widetilde{\eta}^{\mu\nu}\,\nabla_{\mu}t\;\nabla_{\nu}t
=\frac{1+\widetilde{\kappa}_{\mathrm{tr}}}{1-\widetilde{\kappa}_{\mathrm{tr}}}\,,
\end{equation}
which is positive for parameter $\widetilde{\kappa}_{\mathrm{tr}}\in (-1,1]$,
where the value $\widetilde{\kappa}_{\mathrm{tr}}=1$
arises as the limit from below.
As a result, there are no  closed timelike or lightlike curves
of the effective metric, along which
the modified photons could propagate.
This reflects the global causality of the theory considered,
in particular, for the $\widetilde{\kappa}_{\mathrm{tr}}<0$ case
mentioned in the last paragraph of Section~\ref{subsec:Restriction-isotropic}.
See, e.g., Refs.~\cite{Liberati-etal2001,Dubovsky-etal2007}
for further discussion on Lorentz violation and causality.

\setcounter{equation}{0}
\section{Unitarity}
\label{sec:Unitarity-isotropic}
\vspace*{-0mm}

\subsection{Reflection positivity: Simple test}
\label{subsec:Reflection-positivity-general}
\vspace*{-0mm}

Reflection positivity~\cite{OsterwalderSchrader1973,MontvayMunster1994}
is an important property of the Euclidean theory.
It assures, for example, the existence of
an analytic continuation of the Euclidean propagators to Minkowski
propagators, such that the theory in Minkowski spacetime has
a positive semi-definite Hermitian Hamiltonian $\widehat{H}$
and, therefore,
a unitary time evolution operator $\exp(-\mathrm{i}\,\widehat{H}\,t)$.

Following the previous analysis of MCS theory~\cite{AdamKlinkhamer2001},
we restrict the general discussion of reflection positivity
to the special case of reflection positivity of
a Euclidean two-point function. Concretely, reflection positivity of
the Euclidean two-point function corresponds to the following
inequality:
\begin{equation}
\langle 0|\,\Theta\big(\phi(x_4,\mathbf{x})\big)\,\phi(x_4,\mathbf{x})\,|0\rangle
\geq 0\,,
\end{equation}
for a complex  scalar field $\phi(x_4,\mathbf{x})$ in four-dimensional
Euclidean space and the reflection operation
$\Theta$: $\phi(x_4,\mathbf{x})\mapsto \phi^{\dagger}(-x_4,\mathbf{x})$.

With the Fourier decomposition of the scalar field
operator, the following ``weak''
reflection-positivity condition for the scalar Euclidean
propagator $S_E(k_4,\mathbf{k})$ can be
derived~\cite{AdamKlinkhamer2001,OsterwalderSchrader1973,MontvayMunster1994}:
\begin{equation}
S_E(x_4)
\equiv
\int_{\mathbb{R}^3}\mathrm{d}^3k \int^{+\infty}_{-\infty}
\mathrm{d}k_4\,\exp(-\mathrm{i}k_{4}\,x_{4})\,S_E(k_4,\mathbf{k})
\equiv
\int\mathrm{d}^3k\;S_E(x_4,\mathbf{k})\geq 0\,.
\label{eq:reflection-positivity-weak}
\end{equation}
With appropriate smearing functions, also a
``strong'' condition can be derived
\begin{equation}
S_E(x_4,\mathbf{k})\geq 0\,. \label{eq:reflection-positivity-strong}
\end{equation}
The validity of both conditions will be investigated
for the theory considered, isotropic modified Maxwell theory.

\subsection{Reflection positivity and unitarity}
\label{subsec:Reflection-positivity-isotropic}

If the gauge-field propagator is coupled to physical sources,
i.e., a conserved current $j^{\mu}(k)$, then it follows
from current conservation (at the classical level) or the Ward identities
(at the quantum level) that all terms of the propagator
which contain a propagator four-momentum $k_{\mu}$ vanish by contraction
with $j^{\mu}(k)$. Hence, what remains from
the gauge-field propagator \eqref{eq:propagator-result-isotropic}
after projecting on the physical subspace is the first term involving the
metric tensor and the
third term proportional to a bilinear combination
of the fundamental ``four-vector'' $\xi^{\mu}$. Only these two terms describe
the physical degrees of freedom. The pole structure of the
propagator with respect to its momentum is of crucial importance
for unitarity. The relevant pole structure is in the scalar part
\eqref{eq:propagator-result-isotropic-scalar-part} of the propagator.
[The term proportional to $\xi\,\xi$
has an additional pole at $\widetilde{\kappa}_{\mathrm{tr}}=-1$,
which plays no role for our analysis and is excluded anyway
by condition \eqref{eq:kappatildetrace-domain}.]
Hence, it is sufficient to restrict the unitarity analysis to the scalar
part $K$ of the propagator, given by
\eqref{eq:propagator-result-isotropic-scalar-part}.

Since we have shown in Section~\ref{subsec:Wick-rotation}
that Wick rotation is possible, the scalar propagator
part \eqref{eq:propagator-result-isotropic-scalar-part}
is Wick-rotated to Euclidean space.
The resulting Euclidean expression will be denoted by $S_E$.
Recall, that a Wick rotation gives
\begin{equation}
x_4=-\mathrm{i}x^0\,,\quad k_4=-\mathrm{i}k^0\,.
\end{equation}
With our conventions, $S_E(k_4,\mathbf{k})$ is then given by
the negative of the Wick-rotated scalar propagator function:
\begin{align}
S_E(k_4,\mathbf{k})=
S_E(k_4,|\mathbf{k}|)&=
\frac{1}{\left(1+\widetilde{\kappa}_{\mathrm{tr}}\right)
(k_4^2+|\mathbf{k}|^2)-2\,\widetilde{\kappa}_{\mathrm{tr}}\,k_4^2}
 \notag \\
&=\frac{1}{\left(1-\widetilde{\kappa}_{\mathrm{tr}}\right)k_4^2
+\left(1+\widetilde{\kappa}_{\mathrm{tr}}\right)|\mathbf{k}|^2}\,.
\end{align}
In order to show reflection positivity for the scalar part
of the Euclidean propagator, the following expression
needs to be examined:
\begin{equation}
S_E(x_4,|\mathbf{k}|)=\int^{+\infty}_{-\infty}
\mathrm{d}k_4\,\exp(-\mathrm{i}k_4\,x_4)\,S_E(k_4,|\mathbf{k}|)\,.
\label{eq:reflection-positivity}
\end{equation}
Performing the relevant integrals for
$\widetilde{\kappa}_{\mathrm{tr}}$ in the open interval
corresponding to \eqref{eq:kappatildetrace-domain} yields
\begin{subequations}
\begin{equation}\label{eq:S-weak-good}
S_E(x_4,|\mathbf{k}|)=
\frac{\pi}{\sqrt{1-\widetilde{\kappa}_{\mathrm{tr}}^2}}\;
\frac{1}{|\mathbf{k}|}\;\exp\Big(-|x_4|\,\mathcal{B}^{-1}\,|\mathbf{k}|\Big)\,,
\end{equation}
\begin{equation}\label{eq:S-strong-good}
S_E(x_4)=\frac{4\pi^2}{\sqrt{1-\widetilde{\kappa}_{\mathrm{tr}}^2}}\;
\frac{\mathcal{B}^{2}}{x_4^2}\,.
\end{equation}
\end{subequations}
Both of these expressions for $S_E(x_4,|\mathbf{k}|)$ and $S_E(x_4)$ are
manifestly larger than zero for
$\widetilde{\kappa}_{\mathrm{tr}}\in (-1,1]$,
where the value $\widetilde{\kappa}_{\mathrm{tr}}=1$
arises as the limit from below.
Hence, reflection positivity
\eqref{eq:reflection-positivity-weak}--\eqref{eq:reflection-positivity-strong}
is guaranteed for this parameter domain.
In turn, this implies unitarity of the pure-photon sector,
provided $\widetilde{\kappa}_{\mathrm{tr}}\in (-1,1]$.

For $\widetilde{\kappa}_{\mathrm{tr}}\notin [-1,1]$,
the corresponding results are
\begin{subequations}
\begin{equation}
S_E(x_4,|\mathbf{k}|)=
-\frac{\pi}{\sqrt{\widetilde{\kappa}_{\mathrm{tr}}^2-1}}\;
\frac{1}{|\mathbf{k}|}\;\sin\big(|x_4|\,\mathcal{C}\,|\mathbf{k}|\big)\,,
\label{eq:S-strong-bad}
\end{equation}
\begin{equation}
S_E(x_4)=\frac{4\pi^3}{\sqrt{\widetilde{\kappa}_{\mathrm{tr}}^2-1}}\;
\delta'\big(|x_4|\,\mathcal{C}\big)\,,
\label{eq:S-weak-bad}
\end{equation}
with
\begin{equation}
\mathcal{C}\equiv
\sqrt{\frac{\widetilde{\kappa}_{\mathrm{tr}}+1}{\widetilde{\kappa}_{\mathrm{tr}}-1}}\,.
\end{equation}
\end{subequations}
Both of these last expressions for $S_E(x_4,|\mathbf{k}|)$
and $S_E(x_4)$ are not manifestly positive because of the presence
of the sine function and the delta function
derivative.\footnote{Equation~\eqref{eq:S-weak-bad}
is to be understood as acting on a test function.
The sign of the resulting expression depends on the test function
and need not be positive.} As a result, unitarity is
violated for this parameter choice, which is also obvious
from the fact that the corresponding dispersion relation is imaginary.

More generally, it is clear that the pure-photon isotropic
modified Maxwell theory \eqref{eq:L-modMax-isotropic}
is unitary by the following simple argument (prefigured in the
discussion of Section~\ref{subsec:Effective-metric}).
As the electric field involves one derivative
with respect to the spacetime coordinate $x^0\equiv c\,t$,
the Lagrange density \eqref{eq:L-modMax-isotropic}
can be made to be proportional to the standard form
(having $\widetilde{\kappa}_{\mathrm{tr}}=0$)
by the introduction of a rescaled velocity
$c'\equiv c\,\mathcal{B}$, with  constant
$\mathcal{B}\equiv
\sqrt{(1-\widetilde{\kappa}_{\mathrm{tr}})/
      (1+\widetilde{\kappa}_{\mathrm{tr}})}$
as defined in \eqref{eq:dispersion-isotropic}.
Moreover, as the action always appears divided
by the Planck constant $\hbar$, it is even possible
to remove the remaining overall factor by the introduction of a
rescaled constant
$\hbar'\equiv \hbar/(1-\widetilde{\kappa}_{\mathrm{tr}})$.\footnote{Similar
redefinitions bring the commutators \eqref{eq:commutator-ee-eb-bb} back
to the standard Jordan--Pauli form~\cite{JordanPauli1928}.}
Now, standard Maxwell--Jordan--Pauli photons
(even with phase velocity $c'$
and rescaled Planck constant $\hbar'$) have been proven to be
unitary~\cite{Veltman1994,OsterwalderSchrader1973,MontvayMunster1994}.

The outstanding issue is whether or not the unitarity of the
pure-photon isotropic modified Maxwell theory
is affected by the standard minimal
coupling \eqref{eq:standDirac-action} to matter
(recall that this minimal coupling is governed by the gauge principle).
As there is only a single local interaction term
in the action, reflection positivity can be expected to
hold~\cite{MontvayMunster1994}, but this needs, of course, to be verified.

In that respect, it is highly relevant that
two physical decay processes have already been calculated in
Ref.~\cite{KlinkhamerSchreck2008}. The exact tree-level results
for the corresponding decay rates were found to be well-behaved
for parameter values in the domain \eqref{eq:kappatildetrace-domain}.
Clearly, this agrees with the conjecture that isotropic
modified QED \eqref{eq:action-isotropic-modQED} is unitary
for the proper values
of the parameter $\widetilde{\kappa}_{\mathrm{tr}}$.
Further evidence will be given in the next subsection.

\subsection{Optical theorem}
\label{subsec:Optical-theorem-isotropic}

In order to explicitly check the unitarity of isotropic
modified QED \eqref{eq:action-isotropic-modQED} for the parameter
domain $\widetilde{\kappa}_{\mathrm{tr}}\in (-1,1]$, we will consider
one particular process (electron-positron annihilation).
The general idea is that the total cross section
or decay width of a physical process is related to
the imaginary part of the respective forward scattering amplitude via
the optical theorem~\cite{ItzyksonZuber1980,PeskinSchroeder1995}.
The optical theorem follows directly from the unitarity of the
S--matrix. Hence, if unitarity does not hold, this
can be expected to show up as a violation of the optical theorem.

Consider, then, the following (nonstandard)
process involving definite polarization
states of the charged particles:
annihilation of a left-handed electron and a right-handed positron
to a single photon.
The optical theorem will be verified by comparing the imaginary part
of the relevant forward electron-positron
scattering amplitude to the total cross section for
the production of a modified photon ($\widetilde{\gamma}$)
from a left-handed electron ($e^{-}_{L}$) and a
right-handed positron ($e^{+}_{R}$):
\begin{equation}\label{eq:opt-theorem1}  
\hspace*{-5mm}
2\,\mathrm{Im}\left(\begin{array}{c}
\includegraphics{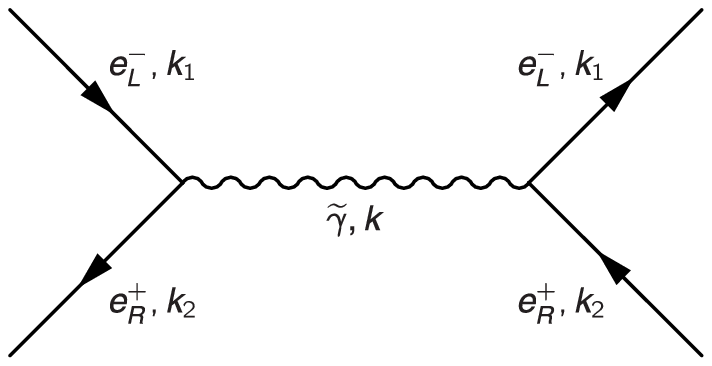}
\end{array}
\right)
\stackrel{?}{=}
\int \mathrm{d}\Pi_1 \left|\begin{array}{c}
\includegraphics{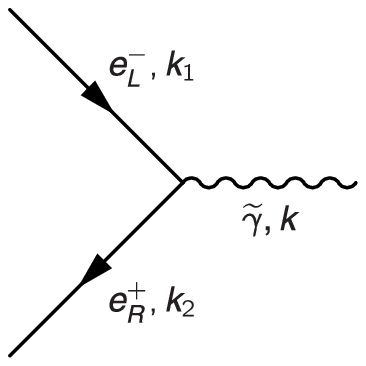}
\end{array}\right|^2\,,
\end{equation}
where $\mathrm{d}\Pi_1$ denotes the one-particle phase-space element of the
modified photon $\widetilde{\gamma}$ in the final state.

Let us take, for simplicity, massless fermions, so that the helicity
of a particle is a physically well-defined property, that is,
independent of the reference frame. (Recall that,
for the case of massive fermions, chirality is not equal to helicity.
In that case, left- and right-handed particles
carry both parallel and antiparallel spins
with respect to the momenta of the particles.)
The assumption of massless particles leads to a
conserved axial vector current:
$\partial_{\mu}\,j_{5}^{\mu}(x)=0$ with
$j_{5}^{\mu}(x)=\overline{\psi}(x)\gamma^{\mu}\gamma_5\psi(x)$.
As a result, we have $k_{\mu}\,j_{5}^{\mu}(k)=0$ for a photon
with momentum $k_{\mu}$ coupling to the current $j_{5}^{\mu}(k)$.
Incidentally, we can neglect the anomalous nonconservation of the
axial vector current~\cite{ItzyksonZuber1980,PeskinSchroeder1995},
because the possible additional terms in our calculation
would be of higher order in the gauge coupling constant $e$.

In the following calculation, it is assumed that
the LV parameter $\widetilde{\kappa}_{\mathrm{tr}}$ is negative and that
the massless particles which annihilate have a nonzero total energy
and are not collinear.
In addition, the chirality conventions
of Ref.~\cite{PeskinSchroeder1995} will be used.
The forward scattering amplitude $\mathcal{M}_1\equiv
\mathcal{M}(e^{-}_{L}e^{+}_{R}\rightarrow
e^{-}_{L}e^{+}_{R})$ is then given by
\begin{align}
\mathcal{M}_1&= e^2\;\overline{u}(k_1)\gamma^{\nu}
\frac{\mathds{1}-\gamma_5}{2}v(k_2)\;
\overline{v}(k_2)\gamma^{\mu}
\frac{\mathds{1}-\gamma_5}{2}u(k_1) \notag \displaybreak[0]\\
&\hspace{3cm}\times\frac{1}{K^{-1}+\mathrm{i}\epsilon}\;
\big(\eta_{\mu\nu}+b\,k_{\mu}k_{\nu}+c\,\xi_{\mu}\xi_{\nu}
+d\,(k_{\mu}\xi_{\nu}+\xi_{\mu}k_{\nu})\big)\,,
\end{align}
for the photon propagator of the isotropic modified  Maxwell theory,
that is, $K$, $b$, $c$, and $d$ taking values from
\eqref{eq:propagator-result-isotropic-scalar-part} and
\eqref{eq:propagator-result-isotropic-coefficients}.

Introducing an
integration over the momentum $k^{\mu}$ of the virtual photon gives
\begin{align}
&\int \frac{\mathrm{d}^4k}{(2\pi)^4}\,\delta^{(4)}(k_1+k_2-k)\,
\mathcal{M}_1
=\notag \displaybreak[0]\\ &\qquad\,
=\int^{+\infty}_{-\infty}
\frac{\mathrm{d}k^0}{2\pi} \int
\frac{\mathrm{d}^3k}{(2\pi)^3}\,\delta^{(4)}(k_1+k_2-k)\;e^2\;
\overline{u}(k_1)\gamma^{\nu}\frac{\mathds{1}-\gamma_5}{2}v(k_2)\;
\overline{v}(k_2)\gamma^{\mu}\frac{\mathds{1}-\gamma_5}{2}u(k_1)
\notag \displaybreak[0] \\ &\hspace{3cm}\,\times \frac{1}{N}\;
\frac{\eta_{\mu\nu}+b\,k_{\mu}k_{\nu}+c\,\xi_{\mu}\xi_{\nu}
+d\,(k_{\mu}\xi_{\nu}+\xi_{\mu}k_{\nu})}
{(k^0-\mathcal{B}\,|\mathbf{k}|+\mathrm{i}\epsilon)
(k^0+\mathcal{B}\,|\mathbf{k}|-\mathrm{i}\epsilon)}\,,
\end{align}
with
\begin{equation}
\frac{1}{N}\equiv\frac{1}{1+\widetilde{\kappa}_{\mathrm{tr}}}\,,
\end{equation}
and $\mathcal{B}$ from  \eqref{eq:dispersion-isotropic}.

For the imaginary part of the amplitude, only the propagator poles
contribute, since Feynman's $\mathrm{i}\epsilon$
prescription only becomes important at the poles.
These poles are given by $k^0=+\mathcal{B}\,|\mathbf{k}|-\mathrm{i}\epsilon$
and
$k^0=-\mathcal{B}\,|\mathbf{k}|+\mathrm{i}\epsilon$,
with a positive infinitesimal
$\epsilon$. The following relation
holds for the propagator pole with a positive real part:
\begin{equation}
\frac{1}{k^0-\mathcal{B}\,|\mathbf{k}|+\mathrm{i}\epsilon}
=\mathcal{P}\frac{1}{k^0-\mathcal{B}\,|\mathbf{k}|}
-\mathrm{i}\pi\,\delta\big(k^0-\mathcal{B}\,|\mathbf{k}|\big)
\equiv\mathcal{P}\frac{1}{k^0-\omega}-\mathrm{i}\pi\,
\delta\big(k^0-\omega\big)\,,
\label{eq:imaginary-part}
\end{equation}
where $\mathcal{P}$ denotes the principal value.
The first term on the far
right-hand-side of \eqref{eq:imaginary-part} is real,
whereas the second one
is imaginary and puts the virtual photon on-shell.
By energy conservation, only the positive-frequency pole is relevant
and, in order to obtain the imaginary part,
all $k^0$ have to be replaced by the frequency $\omega$ from
the dispersion relation \eqref{eq:dispersion-isotropic}.

With $\omega$ defined in \eqref{eq:imaginary-part}
and the further notation $\widehat{\mathcal{M}}_1\equiv
\mathcal{M}(e^{-}_{L}e^{+}_{R}\rightarrow \widetilde{\gamma})$,
we finally get:
\begin{align}\label{eq:optical-theorem-1-final}
2\,\mathrm{Im}(\mathcal{M}_1)&=
-\int^{+\infty}_{-\infty} \mathrm{d}k^0\,
\int \frac{\mathrm{d}^3k}{(2\pi)^3}\,\delta^{(4)}(k_1+k_2-k)
\,\delta(k^0-\omega)
\notag\displaybreak[0] \\
&\hspace{1.5cm}\,\times e^2\;
\overline{u}(k_1)\gamma^{\nu}\frac{\mathds{1} -\gamma_5}{2}v(k_2)\;
\overline{v}(k_2)\gamma^{\mu}\frac{\mathds{1}
-\gamma_5}{2}u(k_1) \notag \displaybreak[0]\\
&\hspace{1.5cm}\,\times \frac{1}{N}\;\frac{\eta_{\mu\nu}
+b\,k_{\mu}k_{\nu}+c\,\xi_{\mu}\xi_{\nu}+d\,(k_{\mu}\xi_{\nu}+\xi_{\mu}k_{\nu})}{k^0+\omega}
\notag \displaybreak[0]\\[1mm]
&=-\int \frac{\mathrm{d}^3k}{(2\pi)^{3}\,2\omega}\,
\delta^{(4)}(k_1+k_2-k) \notag \displaybreak[0]\\
&\hspace{1.5cm}\,\times
e^2\;\overline{u}(k_1)\gamma^{\nu}\frac{\mathds{1}-\gamma_5}{2}v(k_2)\;
\overline{v}(k_2)\gamma^{\mu}\frac{\mathds{1}
-\gamma_5}{2}u(k_1) \notag \displaybreak[0]\\
&\hspace{1.5cm}\,\times \frac{1}{N}\;
\Big(\eta_{\mu\nu}+b\,k_{\mu}k_{\nu}+c\,\xi_{\mu}\xi_{\nu}
+d\,(k_{\mu}\xi_{\nu}+\xi_{\mu}k_{\nu})\Big)
\notag \displaybreak[0]\\[1mm]
&=\int \frac{\mathrm{d}^3k}{(2\pi)^{3}\,2\omega}\,\delta^{(4)}(k_1+k_2-k)
\,(\widehat{\mathcal{M}}_1^{\,\dagger})^{\nu}(\widehat{\mathcal{M}}_1)^{\mu}
\frac{1}{N}\;\Big(-\eta_{\mu\nu}-c\,\xi_{\mu}\xi_{\nu}\Big)
\notag \displaybreak[0]\\[1mm]
&=\int \frac{\mathrm{d}^3k}{(2\pi)^{3}\,2\omega}\,\delta^{(4)}(k_1+k_2-k)\,
(\widehat{\mathcal{M}}_1^{\,\dagger})^{\nu}(\widehat{\mathcal{M}}_1)^{\mu}
\left(\sum_{\lambda=1,\, 2}
\overline{(\varepsilon^{(\lambda)})}_{\nu}
(\varepsilon^{(\lambda)})_{\mu}\right)
\notag \displaybreak[0]\\[1mm]
&=\int
\frac{\mathrm{d}^3k}{(2\pi)^{3}\,2\omega}\,\delta^{(4)}(k_1+k_2-k)
\,\sum_{\lambda=1,\, 2}
|\widehat{\mathcal{M}}_1|^2\,,
\end{align}
where the last line employs
the definition  $\widehat{\mathcal{M}}_1(k)\equiv
\varepsilon_{\mu}(k)\,(\widehat{\mathcal{M}}_1)^{\mu}(k)$.
In the third step of the above derivation,
the Ward identity has been used, so that
all terms vanish for which the momentum
$k_{\mu}$ is contracted with $(\widehat{\mathcal{M}}_1)^{\mu}$
or its Hermitian conjugate. Recall that
the Ward identity~\cite{Veltman1994,ItzyksonZuber1980,PeskinSchroeder1995} reads
\begin{equation}
k_{\mu}\,\mathcal{M}^{\mu}=0\,,
\end{equation}
for a general matrix element $\mathcal{M}^{\mu}(k)$ to which an external
photon [with polarization vector $\varepsilon_{\mu}(k)$ and momentum $k_{\mu}$]
couples. What remains in the fourth
step of \eqref{eq:optical-theorem-1-final} is
the polarization sum \eqref{eq:polarization-sum-isotropic} [or, more precisely,
the truncated polarization
sum \eqref{eq:polarization-sum-isotropic-truncated}],
since $N$ corresponds to the normalization
factor $1/(1+\widetilde{\kappa}_{\mathrm{tr}})$
and $-c$ to
$2\,\widetilde{\kappa}_{\mathrm{tr}}/(1+\widetilde{\kappa}_{\mathrm{tr}})$
according to \eqref{eq:propagator-result-isotropic-c-coeff}.

Two technical remarks on the derivation \eqref{eq:optical-theorem-1-final}
are in order. First, the $k^2=0$ poles in the coefficients $b$ and $d$
from \eqref{eq:propagator-result-isotropic-coefficients}
do not affect the result. Energy-momentum conservation,
encoded by the four-dimensional
$\delta$-function with argument $k_1+k_2-k$,
prevents the pole to be reached
in the integration over $k^0$.
Second, the Ward identity holds not only
for an external photon but also for a photon
propagator coupled to a conserved current.
Hence, the Ward identity could already be used
in the first step of \eqref{eq:optical-theorem-1-final}
to remove the terms with coefficients $b$ and $d$ altogether.
As a result, there would appear no $k^2=0$ poles
in the rest of the calculation.

The conclusion from  \eqref{eq:optical-theorem-1-final} is that
the imaginary part of the forward scattering amplitude of the process
$e^{-}_{L}e^{+}_{R}\rightarrow e^{-}_{L}e^{+}_{R}$
is related to the total cross section for the annihilation process
$e^{-}_{L}e^{+}_{R} \rightarrow \widetilde{\gamma}$.
This result verifies the validity of the optical theorem, at least,
for the process considered.

More generally, it is clear that only the position of
the physical propagator poles and the existence of the Ward identity
are of importance for the validity of the optical theorem
(see, in particular, the succinct discussion of standard QED unitarity in
Chapter~9 of Ref.~\cite{Veltman1994}).
The form of the matrix element itself (whether it is, for example,
polarized or unpolarized) plays no role.

Since both reflection positivity
and the optical theorem have been verified in this section,
we conclude that, most likely, unitarity of the modified QED theory
\eqref{eq:action-isotropic-modQED} holds
for $\widetilde{\kappa}_{\mathrm{tr}}\in (-1,1]$.
The only \textit{caveat} we have is the assumed applicability
of Feynman--Dyson perturbation theory. But perturbation theory appears
to hold for modified QED~\cite{KosteleckyLanePickering2001}, just as
it does for the standard Lorentz-invariant
theory~\cite{Veltman1994,ItzyksonZuber1980,PeskinSchroeder1995}.

\vspace*{-0mm}
\setcounter{equation}{0}
\section{Discussion and outlook}
\label{sec:Discussion-outlook}
\vspace*{-0mm}

In this article, the microcausality and unitarity of the isotropic modified
Maxwell theory \eqref{eq:L-modMax-isotropic} have been established for
numerical values of the ``deformation parameter''
$\widetilde{\kappa}_{\mathrm{tr}}$ lying in the domain
\eqref{eq:kappatildetrace-domain}.\footnote{In
Appendix~\ref{sec:Parity-even-anisotropic},
a similar domain is established
for a particular parity-even anisotropic case of
nonbirefringent modified Maxwell theory.}
In addition, evidence has been presented
that these properties of the pure-photon sector carry over to
the modified QED theory \eqref{eq:action-isotropic-modQED} of photons
minimally coupled to standard Lorentz-invariant Dirac particles. These
results rely on Feynman--Dyson perturbation theory.

The next question is precisely which numerical value of
the $\widetilde{\kappa}_{\mathrm{tr}}$
domain holds experimentally, where $\widetilde{\kappa}_{\mathrm{tr}}=0$
corresponds to exact Lorentz invariance.
Moreover, having a nonzero $\widetilde{\kappa}_{\mathrm{tr}}$
singles out a preferred frame of reference, in which the \emph{Ansatz}
``four-vector'' $\xi^{\mu}$ from \eqref{eq:definition-isotropic-case-xi}
is purely timelike.
We have no idea what the proper reference frame is.
Here, the reference frame is taken to correspond
to the sun-centered celestial equatorial frame (SCCEF).
Another possible choice
would be the frame in which the cosmic microwave background is isotropic.
The strategy is, first, to establish whether or not
$\widetilde{\kappa}_{\mathrm{tr}}$ differs from zero and, then,
to determine the relevant reference frame if the parameter
is indeed nonzero.

Direct laboratory bounds on $|\widetilde{\kappa}_{\mathrm{tr}}|$
in the SCCEF range from the $10^{-2}$ level of
the first experiment~\cite{IvesStilwell1938}
to the $10^{-7}$ and $10^{-8}$ levels of the two most recent
experiments~\cite{Reinhardt-etal2007,Hohensee-etal2010}.
Indirect laboratory bounds are much stronger, ranging from the
$10^{-11}$ level~\cite{Hohensee-etal2008} to the
$5\times 10^{-15}$ level \cite{Altschul2009}.
Still better indirect earth-based bounds follow from the
observation of ultra-high-energy-cosmic-ray (UHECR) primaries
and TeV gamma-rays at the top of the Earth's atmosphere:
$- 0.9 \times 10^{-15}<\widetilde{\kappa}_\text{tr}<0.6\times 10^{-19}$
at the two--$\sigma$ level~\cite{KlinkhamerSchreck2008}.
Future results on UHECRs and TeV gamma-rays may even improve this
last two-sided bound by two orders of magnitude~\cite{Klinkhamer2010}.

The tight experimental bounds on $\widetilde{\kappa}_{\mathrm{tr}}$
can perhaps be understood as implying the extreme smoothness of space,
if $\widetilde{\kappa}_{\mathrm{tr}}$ arises as
the excluded-spacetime-volume fraction of ``defects'' randomly embedded
in flat Minkowski spacetime~\cite{BernadotteKlinkhamer2007}.
Specifically, calculations in simple models give a positive value for
$\widetilde{\kappa}_{\mathrm{tr}}$ proportional to $(b/l)^4$,
where $b$ corresponds to the
typical size of the defect (this size being obtained from measurements
in the ambient flat spacetime) and $l$ to the typical minimal length between the
individual defects (again, from measurements in the ambient flat spacetime).
Remark that, \textit{a priori}, the excluded-spacetime-volume fraction
$(b/l)^4$ can be of order unity, implying the same order of magnitude
for the deformation parameter
$\widetilde{\kappa}_{\mathrm{tr}}$~\cite{BernadotteKlinkhamer2007}.%
\footnote{An alternative
calculation of $\widetilde{\kappa}_{\mathrm{tr}}$ relies on
anomalous effects and gives a positive value proportional to the
product of the fine-structure constant $\alpha$ and the
affected-spacetime-volume fraction from ``punctures'' ($b=0$)
randomly embedded in flat Minkowski spacetime~\cite{KlinkhamerRupp2004}.
Since, \textit{a priori}, the affected--spacetime-volume fraction
can be of order unity, the deformation parameter
$\widetilde{\kappa}_{\mathrm{tr}}$ from anomalous effects
can be of order $\alpha\sim 10^{-2}$.}

This brings us, finally, to the structure of spacetime (and possibly the
cosmological constant problem~\cite{Weinberg1988,KlinkhamerVolovik2009}).
In that respect, it is of
direct relevance that the modified QED theory
\eqref{eq:action-isotropic-modQED} can also be coupled to external
gravitational fields. But, remarkably, modified QED cannot be coupled to
dynamical gravitational fields~\cite{Kostelecky2004,BetschartKK2008}. The
main hurdle appears to be that the energy-momentum tensor $T_{\mu\nu}$ of
isotropic modified QED has an antisymmetric part; see Eq.~(2.11b) of
Ref.~\cite{BetschartKK2008} with $\xi^\mu$ from
\eqref{eq:definition-isotropic-case-xi} in this paper. The conclusion may be
that either standard gravity rules out the particular
theory \eqref{eq:action-isotropic-modQED} with explicit Lorentz
violation or that the theory of gravity itself needs to be modified
fundamentally.

\begin{appendix}

\section{Parity-even anisotropic case}
\label{sec:Parity-even-anisotropic}
\renewcommand{\theequation}{A.\arabic{equation}}\setcounter{equation}{0}
\setcounter{equation}{0}

\subsection{Definition and dispersion relation}
\label{subsec:Definition-parity-even}

One particular anisotropic case of nonbirefringent modified Maxwell theory
\eqref{eq:action-modified-maxwell-theory}--\eqref{eq:nonbirefringent-Ansatz}
is characterized by a  purely spacelike
normalized four-vector $\xi^{\mu}$ in a preferred reference frame
and a single Lorentz-violating
parameter $\widetilde{\kappa}_{33}$:%
\begin{subequations}\label{eq:definition-anisotropic-case}
\begin{eqnarray}
\widetilde{\kappa}^{\mu\nu}&=&
\frac{4}{3}\,\widetilde{\kappa}_{33}
\left(\xi^{\mu}\xi^{\nu}-
\frac{1}{4}\, \xi^{\lambda}\xi_{\lambda}\, \eta^{\mu\nu}
\right)\,,\\[2mm]
(\xi^{\mu})&=&(0,\,0,\,0,\,1)\,,\\[2mm]
(\widetilde{\kappa}^{\mu\nu})&=&
\widetilde{\kappa}_{33}\;
\mathrm{diag}\left(\frac{1}{3},-\frac{1}{3},-\frac{1}{3},\,1\right)\,.
\label{eq:definition-anisotropic-case-widetildekappamunu}
\end{eqnarray}
\end{subequations}
By choosing the momentum four-vector as
\begin{equation}
(k^{\mu})=\big(\omega(\mathbf{k}),\,k_{\bot},\,0,\,k_{\|}\big)\,,\quad
k_{\|}=\mathbf{k}\cdot \boldsymbol{\xi}\,,\quad
k_{\bot}=|\mathbf{k}-k_{\|}\,\boldsymbol{\xi}|\,,\quad
\boldsymbol{\xi}=(0,\,0,\,1)\,,
\end{equation}
we obtain the following dispersion relation from the field equations
\eqref{eq:field-equations-modified-maxwell-theory}:
\begin{equation}\label{eq:disp-rel--parity-even}
\omega(\mathbf{k})=\sqrt{k_{\bot}^2+\mathcal{D}^2\,k_{\|}^2}\,,\quad
\mathcal{D}\equiv\sqrt{\frac{1-2\,\widetilde{\kappa}_{33}/3}{1+2\,\widetilde{\kappa}_{33}/3}}\,.
\end{equation}

The case considered can be expressed in terms of the
standard-model-extension (SME) parameters
\cite{KosteleckyMewes2002} with the help of the
``translation dictionary'' from \cite{KlinkhamerRisse2008b}:
\begin{equation}
\widetilde{\kappa}_{\mathrm{tr}}=
\frac{2}{9}\;\widetilde{\kappa}_{33}\,,\quad
(\widetilde{\kappa}_{\mathrm{e-}})^{(11)}=
\frac{4}{9}\;\widetilde{\kappa}_{33}\,,\quad
(\widetilde{\kappa}_{\mathrm{e-}})^{(22)}=
\frac{4}{9}\;\widetilde{\kappa}_{33}\,.
\end{equation}
Hence, the anisotropic case considered in this appendix
involves a mixture of the three parity-even parameters
$(\widetilde{\kappa}_{\mathrm{e-}})^{(11)}$,
$(\widetilde{\kappa}_{\mathrm{e-}})^{(22)}$,
and $\widetilde{\kappa}_{\mathrm{tr}}$.

\subsection{Microcausality}
\label{subsec:Microcausality-parity-even}

The commutators of electric and magnetic fields can be computed just as
for the  isotropic case in Section~\ref{subsec:Commutator-gauge-potential}.
The results involve a particular tensor structure
and a scalar commutator function (here, distinguished by a bar)
\begin{equation}
\overline{D}(x)=
\oint_{C} \frac{\mathrm{d}k_0}{2\pi} \int
\frac{\mathrm{d}^3k}{(2\pi)^3}\,
\frac{1}{\left(1+2\,\widetilde{\kappa}_{33}/3\right)
(k_0^2-k_{\bot}^2)-\left(1-2\,\widetilde{\kappa}_{33}/3\right)k_{\|}^2}
\,\exp\big(\mathrm{i}\,k_0\,x_0+\mathrm{i}\,\mathbf{k}\cdot\mathbf{x}\big)\,.
\label{eq:commutator-function-configuration-space-anisotropic-case}
\end{equation}
For the issue of microcausality, the properties of
$\overline{D}(x)$ are important and
the computation of the four-dimensional integral in
\eqref{eq:commutator-function-configuration-space-anisotropic-case} yields:
\begin{equation}
\overline{D}(x)=-\frac{1}{2\pi\sqrt{1-4\,\widetilde{\kappa}_{33}^2/9}}\;
\mathrm{sgn}(x_0)\,
\delta\left(x_0^2-x_{\bot}^2-x_{\|}^2\,/\,\mathcal{D}^2\right)\,.
\end{equation}
Hence, analogously to the isotropic case of
Section~\ref{subsec:Commutator-gauge-potential}, the commutator function
vanishes everywhere except on the modified null-cone,
\begin{equation}
x_0^2-x_{\bot}^2-x_{\|}^2\,/\,\mathcal{D}^2=0\,.
\end{equation}
As a result, microcausality is a property also for this particular
anisotropic case of nonbirefringent modified Maxwell theory, provided
\begin{equation}\label{eq:kappatilde33-domain}
\frac{2}{3}\,\widetilde{\kappa}_{33}\in I\,,\quad I\equiv (-1,1]\,,
\end{equation}
which matches the domain of the
isotropic parameter \eqref{eq:kappatildetrace-domain}.
In fact, the formal structure of these two cases is
similar --- recall the definitions of
the matrices $\widetilde{\kappa}^{\mu\nu}$ in
\eqref{eq:definition-isotropic-case-widetildekappamunu} and
\eqref{eq:definition-anisotropic-case-widetildekappamunu}.
Note also that these two cases of nonbirefringent modified
Maxwell theory, for $\widetilde{\kappa}_{\mathrm{tr}}>0$
and $\widetilde{\kappa}_{33}>0$, can be induced from a single
Lorentz-violating term in the fermionic action~\cite{Gomes-etal2009}.

\subsection{Reflection positivity and unitarity}
\label{subsec:Reflection-positivity-parity-even}

The simple test of reflection positivity from
Section~\ref{subsec:Reflection-positivity-general}
works just as for the  isotropic case, since the scalar part of the
Euclidean propagator (here, distinguished by a bar) is
\begin{align}
\overline{S}_E(k_4,\mathbf{k})
&=\frac{1}{\left(1+2\,\widetilde{\kappa}_{33}/3\right)
(k_4^2+|\mathbf{k}|^2)-(4\,\widetilde{\kappa}_{33}/3)k_3^2}
 \notag \\
&=\frac{1}{1+2\,\widetilde{\kappa}_{33}/3}\;
\frac{1}{k_4^2+k_{\bot}^2+\mathcal{D}^2\,k_{\|}^2}\,.
\end{align}

Turning immediately to the strong reflection-positivity
condition \eqref{eq:reflection-positivity-strong} for a deformation
parameter $\widetilde{\kappa}_{33}$ from \eqref{eq:kappatilde33-domain},
the calculation of the integral gives:
\begin{align}\label{eq:reflection-positivity-strong-parity-even}
\overline{S}_E(x_4,\mathbf{k})&=\int_{-\infty}^{+\infty}
\mathrm{d}k_4\,\exp(\mathrm{i}\,k_{4}\,x_{4})\,\overline{S}_E(k_4,\mathbf{k})=
\frac{1}{1+2\,\widetilde{\kappa}_{33}/3}\int_{-\infty}^{+\infty}
\mathrm{d}k_4\;\frac{\exp(\mathrm{i}\,k_{4}\,x_{4})}{k_4^2+\omega(\mathbf{k})^2}
\notag \\
&=\frac{2}{1+2\,\widetilde{\kappa}_{33}/3}\int_{0}^{\infty}
\mathrm{d}k_4\;\frac{\cos(k_{4}\,x_{4})}{k_4^2+\omega(\mathbf{k})^2}=
\frac{1}{1+2\,\widetilde{\kappa}_{33}/3}\;\frac{\pi}{\omega(\mathbf{k})}\;
\exp\big(-|x_4|\,\omega(\mathbf{k})\big)\,,
\end{align}
with $\omega(\mathbf{k})$ given by \eqref{eq:disp-rel--parity-even}.
Result \eqref{eq:reflection-positivity-strong-parity-even},
for $\widetilde{\kappa}_{33}$ given by \eqref{eq:kappatilde33-domain},
proves the strong reflection-positivity condition for
the scalar propagator. Based on the result for the isotropic case,
unitarity can be expected to hold also for
the particular case of anisotropic modified Maxwell theory
\eqref{eq:definition-anisotropic-case} interacting with a standard matter
sector \eqref{eq:standDirac-action}.

\subsection{Discussion}
\label{subsec:Discussion-parity-even}

As shown in this appendix, the pure-photon sector of the parity-even
anisotropic nonbirefringent modified Maxwell theory characterized by
parameters \eqref{eq:definition-anisotropic-case}
has microcausality and unitarity for the
$\widetilde{\kappa}_{33}$ parameter
domain \eqref{eq:kappatilde33-domain}.
The same can be expected to hold for the modified QED theory
of photons minimally coupled to standard Lorentz-invariant
Dirac particles \eqref{eq:standDirac-action}.
The question, now, is which numerical value
of $\widetilde{\kappa}_{33}$
holds experimentally, where $\widetilde{\kappa}_{33}=0$
corresponds to having exact Lorentz invariance.

The isolated Lorentz-violating parameters
$(\widetilde{\kappa}_{\mathrm{e-}})^{(11)}$ and
$(\widetilde{\kappa}_{\mathrm{e-}})^{(22)}$
are already tightly bounded at the $10^{-17}$ level by direct laboratory
experiments~\cite{Hohensee-etal2010,Eisele-etal2009,Herrmann-etal2009}.
This implies that, at the present moment,
the experimental limit on the $\widetilde{\kappa}_{33}$
parameter of the particular case considered in this appendix is controlled
by the less tight limits on $\widetilde{\kappa}_{\mathrm{tr}}$,
which have been discussed in the
third paragraph of Section~\ref{sec:Discussion-outlook}.

\end{appendix}



\end{document}